# DELINEATION OF TECHNIQUES TO IMPLEMENT ON THE ENHANCED PROPOSED MODEL USING DATA MINING FOR PROTEIN SEQUENCE CLASSIFICATION

Ananya Basu[1] and Suprativ Saha[2]

[1]Department of Computer Science and Engineering,
Global Institute of Management and Technology, India
[2]Department of Computer Science and Engineering,
Adamas Institute of Technology, India

**ABSTRACT**

*In post genomic era with the advent of new technologies a huge amount of complex molecular data are generated with high throughput. The management of this biological data is definitely a challenging task due to complexity and heterogeneity of data for discovering new knowledge. Issues like managing noisy and incomplete data are needed to be dealt with. Use of data mining in biological domain has made its inventory success. Discovering new knowledge from the biological data is a major challenge in data mining technique. The novelty of the proposed model is its combined use of intelligent techniques to classify the protein sequence faster and efficiently. Use of FFT, fuzzy classifier, String weighted algorithm, gram encoding method, neural network model and rough set classifier in a single model and in an appropriate place can enhance the quality of the classification system .Thus the primary challenge is to identify and classify the large protein sequences in a very fast and easy but intellectual way to decrease the time complexity and space complexity.*

## 1. INTRODUCTION

As the world has emerged into a new era of technology, data have become boundless especially when the technologies like computer, mass storage media etc. has led to collection of large amount of data. It's a very well-known fact that data mining has taken over the traditional system of analysis of data as the traditional system is not as efficient as data mining technique in handling the large amount of data as well as the noisy data and processing them. Extraction of knowledge from large amount of data is what data mining refers to. The main step in this process is known as Knowledge Discovery in Databases (KDD), although the two terms are often used interchangeably. It is an interdisciplinary area of research that has its roots in databases, machine learning, and statistics. Data mining has been implemented in many other fields of research like pattern recognition, visualization and many more. Biological data has also been increasing at a high rate with the help of new technologies. These data can be analyzed and new information can be obtained from past which can help to predict the future diversities. Some of the most popular tasks are classification, clustering, association and sequence analysis, and regression. Depending on the nature of the data as well as the desired knowledge there is large number of algorithms for each task. All of these algorithms try to fit in a model. Such a model can be either predictive or descriptive. A predictive model makes a prediction about data using known examples, while a

DOI : 10.5121/ijdms.2014.6105     61

descriptive model identifies patterns or relationships in the data. Classification is the most important technique to identify a particular character or a group of them. This paper is based on classification of protein sequences. Popular technique for protein sequence classification is extraction of specific features from the protein sequence. These features depend on the structural and functional properties of amino acid. Researchers apply some well-known classification technique like neural networks, Genetic algorithm, FUZZY ARTMAP, ROUGH SET Classifier etc. Till date none have achieved 100% accuracy level. This paper presents a literature survey of the on-going research on protein sequence classification in section 2, a comparative study of the various classifications in section 3, followed by a new proposed model in section 4 finally the result analysis in section 5 and last but not the least the conclusion in section 6.

## 2. CURRENT STATE OF ART

Different classification techniques have been used to classify protein sequence into its particular class, sub class or family. All these methods aim to extract some features, match the value of these features and finally classify the protein sequence.

### 2.1 Neural Network Model

The vital technique of protein classification "the neural network" [1] [2] [3] [4] [8] [10] is advantageous over many other approaches as the extracted features are distributed in high dimensional space having complex behaviors. Neural network based classifier is proved to be more worthy than tree based techniques to classify protein sequence. Implementation of this particular technique is done by working on the extracted feature using methods like 2-gram encoding method and 6-letter exchange method from the input protein sequence. Formation of pattern matrix including those features is constructed.

Normally neural network is good at handling non-linear data (noise data). But as the protein sequence is linear, neural network does not help much. The data being linear in classification of protein, the use of neural network modeling fails to add any much benefit. Thus limiting the accuracy, neural network model provide 90%-92% accuracy. Improvement of this accuracy is mostly needed.

### 2.2 Fuzzy ARTMAP Model

Fuzzy ARTMAP [6] [7] [9] mode is one of the important among many other methods to classify the protein sequence. The fuzzy model mainly calculates membership value for each and every feature using membership functions. Fuzzy model differs from the neural network by analyzing the data individually. This model was constructed to classify the unknown protein sequence into different predefined protein families or protein sub families which had 93% high accuracy. The molecular weight (W) and the isoelectric point (pI) of the protein sequences followed by Amino acid composition of the sequences were calculated. The Hydropathy composition (C), the Hydropathy distribution (D) and the Hydropathy transmission (T) were also calculated. After extraction of all forty different features an unknown protein sequence was used as the input of the Fuzzy ARTMAP model. But with many advantages there are many disadvantages .This modeling helps in the data analysis although storage and time requirement are high. The construction of fuzzy sets, for every iteration increases the computational complexity as well. This model also failed to process the physical relationships which are most important in this purpose.

### 2.3 Rough Set Classifier

Rough set classifier [11] [12] is capable of overcoming the disadvantages of above classifiers. The large numbers of extra feature, extracted for rule discovery makes the classifiers work more complicatedly which sometimes are unable to handle. As a result they try to select the features to reduce the computation time. But these methods also degrade their performance. Due to feature selection, the accuracy level reduces as each and every feature is important for a fine and smooth classification. Rough set classifier is a new model to overcome this problem. Rough sets theory is a component of hybrid solutions of machine learning method and data mining. The indiscernibility relation that induced minimal decision rules from training examples is the important notation in rough set model. The identification of minimal set of features is done by if–else rule on the decision table. A new approach to compute predominant attributes (approximate reduce) and use them to construct decision tree called Reduce based Decision Tree (RDT) is proposed. Decision rules generated from the RDT are stored in RDT Rules Database (RDTRD). These rules are used to obtain class information. The infirmity of RDT is overcome by extracting spatial information by means of Neighborhood Analysis (NA). Spatial information is converted into binary information using threshold. It is utilized for the construction of Concept Lattice (CL). The Rough Set Classifier model provides knowledge based information only without any analysis of data. For properly classifying the protein sequences, both play an important role. Instead of classifying protein sequence into classes or sub classes, this model provides a small known sequence from a long unknown protein sequence. Thus it requires extra time and space for further classification of the output sequence into classes or sub classes. The accuracy level is 97.7%, which leaves scope for improvement.

## 2.4 Application of Fast Fourier Transformation on Protein Sequence Classification

It [13] provides a spectral domain features for the prediction of protein sequence using SCOP classification. For each protein sequence in the database molecular weight is calculated. Fast Fourier transform is then calculated for every sequence as good discriminating feature. Then neural network classifier is used on the base of the features. The phases are extraction of features based on FFT of molecular weight of each protein sequence, feature selection, neural network based classification. This paper provides a new set of spectral domain features for the prediction of protein classification.

## 2.5 A new Classifier including Neural Network, Fuzzy and Rough Set Classifier

In [14] [15] [16] respect of failure of traditional data analysis technique to process on large amount of data, data mining was invented. Basically Data mining technology can extract knowledge from a large database (KDD) with the help of computer system. Collection of biological data like protein sequence, DNA sequence is increasing at a large number. So as a result of failure of the traditional data analysis techniques, data mining is used to extract meaningful information from a huge amount of biological data and classifies into classes or families.

## 2.6 String Weighted Schema

In this classifier [5] a discriminative approach and particular support vector machines for protein classification is used .String weighting scheme approach based on HMM is used. HMM model is used for extracting features from protein sequence to find maximum sequence information within the features generated. Algorithm for string weighting scheme method is that all protein sequences are converted to high dimensional features through HMM. After this transformation SVM discriminator separates each protein family from the rest. Feature vector generation process converts the protein sequence to short subsequent string of fixed length. If two sequences contain many of the same k-length subsequently, their HMM scores are similar or close to each other. SVM has high efficiency i.e. they do not require any complex tuning of parameters The technique focuses on the conserved and non-conserved regions of proteins of interest, but for conserved region, the score is very low and if the sequence contain much of the region, it makes it hard for the SVM to achieve the good accuracy.

International Journal of Database Management Systems ( IJDMS ) Vol.6, No.1, February 2014## 3. COMPARATIVE ANALYSIS

| Techniques | Database Uses | Features Selection | Accuracy Level | Drawbacks |
|---|---|---|---|---|
| **Neural Network based Classifier** [1, 2, 3, 4, 8, 10] | The Int. Protein Seq. Database Release 62 | • 2-gram encoding method<br>• 6-letter exchange group methods.<br>• Len, Mut, and occur calculation.<br>• Min. description length (MDL) principle | 90% | • Better for Non-linear and Noisy data.<br>• Does not handle Physical relationship. |
| **Fuzzy ARTMAP based Classifier** [6, 7, 9] | SCOP 1.69 ASTRAL1.69 | • Molecular weight<br>• Isoelectric point<br>• Hydropathy composition<br>• Hydropathy distribution<br>• Hydropathy transmission | 93% | • Concerned only about the physical Structure of AA.<br>• Does not handle Physical relationship. |
| **Rough Set based Classifier** [11, 12] | NCBI (Blast) | • Sequence Arithmetic<br>• Reduce based Decision Tree<br>• Neighbourhood Analysis<br>• Concept Lattice | 97.7% | • No analytical output.<br>• Need Extra Time and Space |
| **FFT based Classifier** [13] | NCBI | • 2-gram encoding method<br>• Fast Fourier Transform | 92 % | • Does not handle Physical relationship<br>• Accuracy Level is low but takes more time |
| **Classifier including Neural Network, Fuzzy and Rough Set Classifier** [14, 15, 16] | SCOP 1.69 ASTRAL1.69 NCBI (Blast) | • 2-gram encoding method<br>• 6-letter exchange group methods.<br>• Molecular weight<br>• Isoelectric point<br>• Hydropathy composition<br>• Hydropathy distribution<br>• Hydropathy transmission<br>• Neighbourhood Analysis | 98.5% | • Accuracy level need to improve<br>• Noise removal algorithm is not so strong |
| **String Weighted Schema based Classifier**[5] | NCBI | • 2-gram encoding method<br>• Molecular weight (W)<br>• String Weighted Schema | 93% | • Takes huge time to implement string weighted schema<br>• Accuracy level needs to be improve. |

64

## 4. PROPOSED MODEL

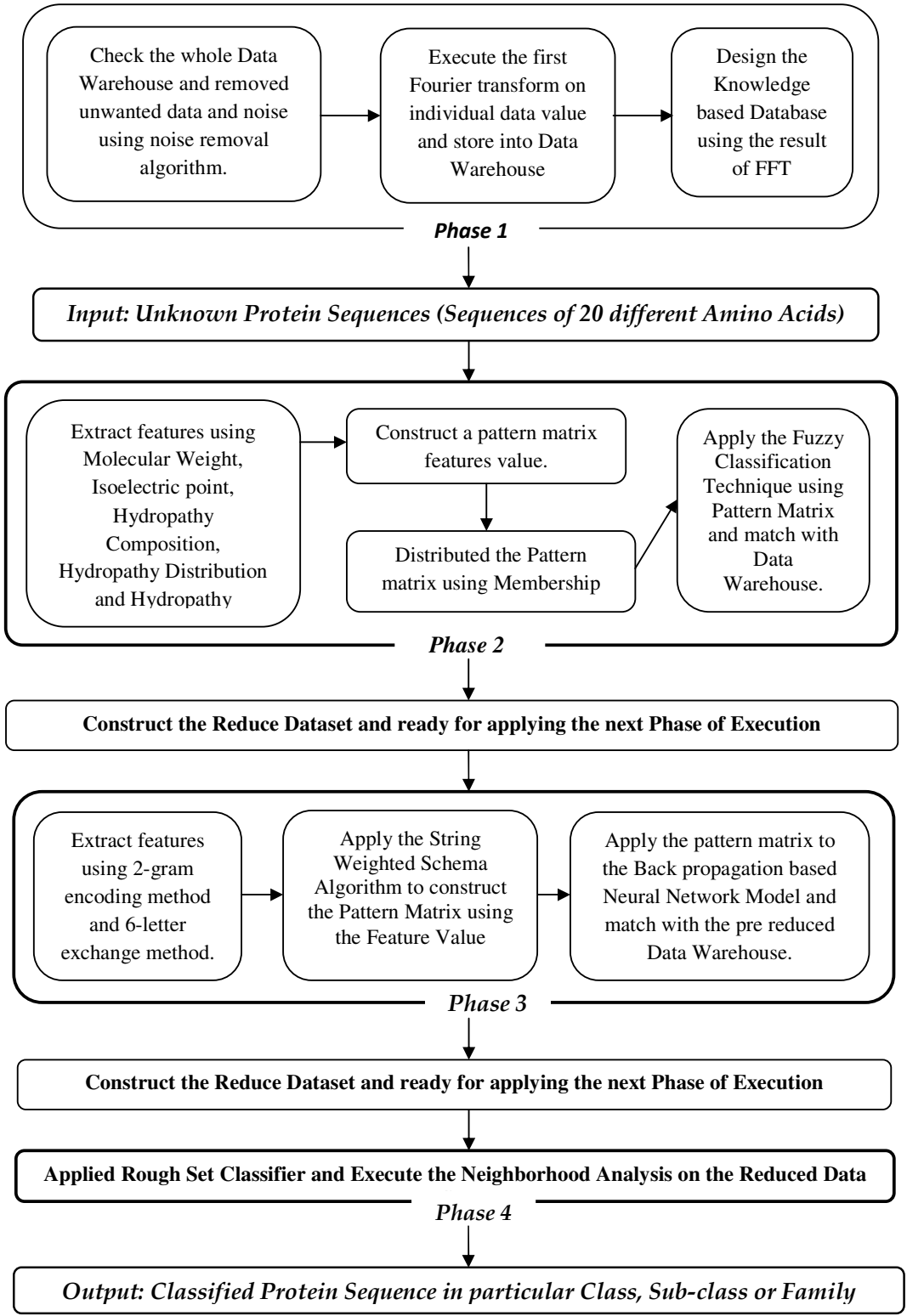

Fig 1: Proposed Model

The purpose of this model is to classify the unknown protein sequence in to different families, with high accuracy and low computational time. To implement it, unknown protein sequence are taken as an input and extract some different features in different phases and match with predefine features values with respect to protein family to classify the sequence into classes or families. First of all, extraction of features is most important. It is an intellectual fact that from the first level any phase can classify the unknown protein sequence in to its family without the execution of the next phase. In this way complexity of this algorithm can be reduced.

In **Phase one**, before the intake of the protein sequence, noise and unwanted data are reduced by using noise removal algorithm on the given data warehouse. This will shorten the database from unwanted data. Then Fast Fourier Transform is applied on the individual feature value stored in the data warehouse related to this problem. At the final stage of phase 1 the knowledge is discovered from database using the result of the FFT [13]. Using FFT on the feature values, the range of the particular feature on individual family can be reduced so that it will be very easy to compare to the input. It's true that, time complexity of the algorithm little bit increases in this phase but in this way accuracy level can be increased more and more. The time complexity which increases here can be solved in the next phases.

In **Phase two**, the input sequence of unknown protein sequence is taken and the features like the molecular weight, isoelectric point, Hydropathy Composition, Hydropathy Distribution and Hydropathy Transmission of the sequence are extracted. Then a pattern matrix using the feature value is formed. Finally the fuzzy ARTMAP [6] [7] [9] model is applied and matched with the knowledge database. It is possible that phase 2 can provide the final result i.e. input unknown protein sequences can be classified using only Fuzzy ARTMAP model. Otherwise Fuzzy model decreases the dataset by matching the data warehouse data by data. In this way time complexity can easily be reduced without changing the accuracy level.

In **Phase three** the features are refilled by extracting them using 2-gram and 6 letter exchange method [1] [2] [8]. The construction of pattern matrix is done using the feature value with the help of string weighted schema. This SVM [5] discriminator separates each protein family from the rest. The string is converted into short subsequent fixed length i.e. Protein sequence of any length is converted into strings and then to any feature vector of fixed length. Next the matrix is applied based on back propagation process of Neural Network Model. The reduced data set is collected.

In the **final Phase**, Rough set classifier is applied and Neighborhood Analysis [11] [12] is used on the reduced dataset to classify the rest of the protein sequences.

Protein analysis has a major role in the part of biological domain .This proposed model could be used as a technique to analyze protein sequences and classify them into groups and sub groups specially classes and families in a more integrated way. The proposed model shall minimize the time variation in computing the result and provide more accuracy in the result. Not only this, the model shall be able to be implemented on large protein sequences of bigger length as an input. The use of FUZZY MODEL can bring accuracy by extracting out the features one by one. Use of FFT technique before the fuzzy model can reduce the dataset at the beginning of the implementation which can make it easy for the rest of the process to go in a very compact way with consuming less time.

## 5. RESULT ANALYSIS

The proposed model is basically tested on 5 protein families. Results of all feature values which are described in the proposed model of the at least 500 protein sequences of each family are stored in the data warehouse to test the proposed model. The computational time is far better than the previous established model related to this field. It is right that noise removal algorithm takes more time than the others but it also reduce the overall data set from the data warehouse so that the final computational time was comparatively better than others. Now the important part is accuracy level analysis. At first around 300 known protein sequences (Data collects form NCBI) with in this five families are measured by this proposed model. It is highly appreciable; within 300 protein sequences all sequences provide the accurate results. Finally tested this model with unknown protein sequences if the protein name is present in our dataset provides 99% accuracy. The use of various database like SCOP, NCBI even adds some help to the work We find the accuracy level of Neural Network classification is 90%.For the Fuzzy model classification it is 93%.Rough set classifier gives an accuracy of 97.7 % whereas the FFT shows 92% accuracy ,the string weighted classifier has 93% accuracy and finally the combined model of all the all the model shows the accuracy of 98.2% which is higher than all the classifiers around 99.12%.

Overall testing is done by own designed tool, which is decorated by JAVA. This tool can identify the unknown protein family; insert new data in the data warehouse and it also have some capability to calculate the computational time of total process.

## 6. CONCLUSION

This dissertation includes a detail review of research work involving six different techniques to classify the protein sequences. It has been observed that knowledge based and analysis of data form integral parts of protein sequence classification. From the computation of this techniques we can come to the conclusion that an increase in accuracy level is seen. The accuracy level of each proposed model has been studied. Finally, a new classification model is proposed and computed which can classify the unknown protein sequences into families, classes or sub classes, producing knowledge based information beside data analysis technique. Designed proposed model is better than others in respect to the computational time and accuracy level. In case of the special characteristics of biological data, the variety of new problems and the extremely high importance of bioinformatics research a large number of critical issues are still open and demands active and collaborative research by the academia as well as the industry. More over new technologies such as the micro areas led to a constantly increasing number of new questions on new data .The technique uses very modified and fast technologies like FFT, FUZZY MODEL, SVM, and Neural Network along with Noise removal Algorithm and accordingly using more intelligence the model can be later modified for better computation. In future different analysis will be done with this new classification model.

International Journal of Database Management Systems ( IJDMS ) Vol.6, No.1, February 2014

## AUTHORS

**Ananya Basu** is Pursuing B-Tech in Computer Science and Engineering from Global Institute of Management and Technology. Her research interests include the field of Database Management System, Data Mining and Distributed Database. Her area of research is Biological Data Mining.

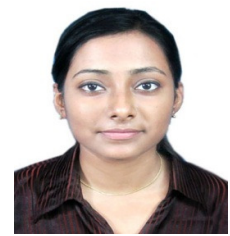

**Suprativ Saha** is an Assistant Professor in Department of Computer Science and Engineering at Adamas Institute of Technology, West Bengal India. Since January 2014. Previously He was associated with Global Institute of Management and Technology and West Bengal University of Technology. He received his M.E degree in year 2012 from West Bengal University of Technology. He is a life member of Operational Research Society of India, Kolkata Chapter. His research interests include the field of Database Management System, Data Mining and Distributed Database. His area of research is Biological Data Mining. Mr. Saha has about 4 referred international and national publications to his credit.

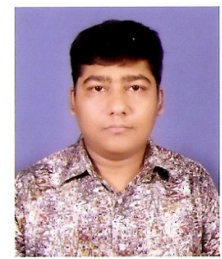